# Surface-emitting electroholographic SAW modulator

Joy C. Perkinson, Michael G. Moebius, Elizabeth J. Brundage, William A. Teynor, Steven J. Byrnes, James C. Hsiao, William D. Sawyer, Dennis M. Callahan, Ian W. Frank[1], John J. LeBlanc, and Gregg E. Favalora[*]

*The Charles Stark Draper Laboratory, Inc., 555 Technology Square, Cambridge, MA 02139, USA*
[1]*Employee of Draper when the reported work was performed*
*\*Corresponding author: gfavalora@draper.com*

**We report the design and operation of a surface-emitting surface acoustic wave (SAW) acousto-optical modulator which behaves as a cm-scale linear hologram in response to an applied electronic waveform. The modulator is formed by an optical waveguide, transducer, and out-coupling surface grating on a 1 mm-thick lithium niobate substrate. We demonstrate the ability to load and illuminate a 9-region linear hologram into the modulator's 8 mm-long interaction region using applied waveforms of 280–320 MHz. To the best of the authors' knowledge, this is the first demonstration of a monolithically-integrated, surface-emitting SAW modulator fabricated using lithographic techniques. Applications include practical implementations of a holographic display.**

Electroholographic three-dimensional (3-D) display technologies rely principally on diffractive phenomena to project distributions of electromagnetic radiation, and are hoped to offer the ultimate expression of synthetic realism [1]. However, canonical hypothetical autostereoscopic applications, such as interventional medical imaging, terrain visualization, and geophysics, still lack an electroholographic display with the display area, image fidelity, and compact packaging of mature 2-D display products. This is primarily due to the need for improved light modulators [1–4].

Approaches to electroholography include pixelated electrically- or optically-addressed spatial light modulators (SLMs), acousto-optical modulators (AOMs), and systems relying on photorefractive polymers [2,4]. Pixelated SLMs, the most prevalent approach, usually have $cm^2$-scale areas, relatively wide package borders, and pixels larger than visible wavelengths. These attributes result in displays which trade off area, viewing angle, frame rate, and package size [5]. AOMs, which convert electronic waveforms into regions of diffraction in the bulk or at the surface of a piezoelectric crystal, have shown promise for 3-D in several forms. Multi-channel bulk-mode AOMs in a descanned Scophony arrangement generate interactive holograms but require electromechanical scanners and large demagnification optics [6,7]. Surface acoustic wave (SAW) AOMs [8,9] exploit piezoelectrically-induced surface waves and are more easily arrayed and can offer higher bandwidth than bulk AOMs for 3-D display [10,11]. In *leaky-mode* SAW AOMs, an applied electronic waveform creates a SAW which interacts with waveguided light in an *interaction region*, causing that light to "leak" a polarization-rotated optical signal into the modulator substrate bulk at angles corresponding to the waveform.

Two arrangements of leaky-mode SAW AOMs (hereinafter simply "SAW AOMs") of particular promise for electroholographic displays are *edge-emitting* [11,12], in which the length of the modulator is sufficiently short for the diffracted light to exit a substrate edge, and an emerging class of *surface-emitting* SAW AOMs, as in Jolly, et al. [13] and our reported device of Fig. 1a. We describe their operation after a summary of their capabilities.

Surface-emitting SAW AOMs exhibit benefits of particular relevance to future handheld or desktop 3-D displays. First, the pixel pitch can be set by choice of fabricated waveguide spacing ($z$ direction in Fig. 1), such as 0.02 mm, 0.1 mm, 1 mm, etc., and SAW waveform design along $y$, which is a continuous-time signal. A second benefit is the utility of that continuous linear holographic modulation along the modulator $y$ axis to support a variety of fringe codings from the field of computational holography, such as holographic elements (*hogels*), small regions of "homogeneous [diffractive] spectrum" [14], or wavefront-curving *wafels* [15]. As a waveform is applied to the device, a linear hologram of appreciable length (order of cm, depending e.g. on SAW attenuation during propagation) can be partitioned. Third, due to this, the number of IDTs and in-coupling ports is greatly reduced compared to edge-emitting SAW AOMs. For example, this Letter reports on producing the equivalent of 9 hogels from a single IDT and illuminator rather than 9 of each in a hypothetical edge-emitting equivalent. The extension of this technology to specific holographic display implementations, such as through hogel [14] or wafel [15]-based approaches, is outside the scope of this report.

Jolly et al. [13,16] report progress towards, but not device-scale demonstration of, surface-emitting modulators fabricated with laser micromachining techniques that use volume hologram out-coupling features. Alternatively, McLaughlin et al. [17] of BYU fabricated and tested an arrangement whose out-coupler is a surface relief grating and directs light from an edge-emitting device into a surface-emitting device. In contrast to these, ours is a single-die modulator implementation compatible with mass-production techniques.

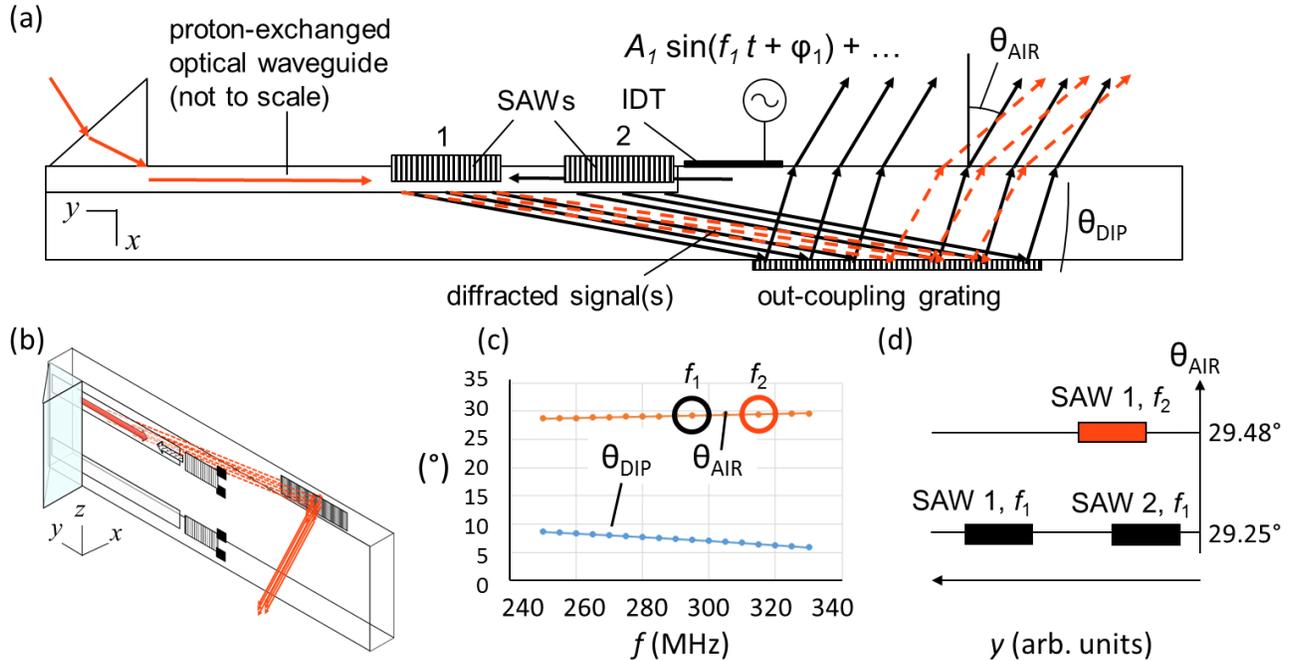

Fig. 1. (a) Side view of surface-emitting SAW AOM. TE light is in-coupled into the waveguide via a rutile prism (not to scale), and interacts with counter-propagating SAW pulses which cause TM light to "leak" into the substrate bulk at angle $\theta_{DIP}$. A SAW typically penetrates one acoustic wavelength (approximately 10 μm in these devices) into the substrate, enabling an interaction with the optical wave confined to a surface waveguide of similar depth. The diffracted light is redirected towards the top modulator face by a 360 nm surface grating and exits at angle $\theta_{AIR}$. In this example, a two-frequency-component SAW 1 yields diffracted optical signals depicted in black and dashed orange, and SAW 2 has one frequency component whose corresponding diffracted signal is depicted in black. (b) Isometric view. (c) Expected dip and exit signal trajectories for various values of $f$. (d) Modulated light exits the surface from a location along $y$ as a function of the location and frequency spectrum of the illuminated SAW. These can be plotted in an angle-space parameterization.

In this Letter, we report to the authors' knowledge the first demonstration of a monolithically-integrated, surface-emitting SAW modulator fabricated using lithographic techniques. In contrast to [13,16], it uses a backside, rather than volume, out-coupling grating, and in comparison to [17] does so in a single prism-coupled integrated optical device. To illustrate a reduction of required RF and optical inputs, 8 mm of electronically-defined diffracting regions (also referred to here as hogels or SAW bursts) were loaded into a single electrode on the device and illuminated with a single beam. Various partitions of the linear holograms are tested, ranging from a single 1 mm-scale diffractive beam-steering region to a group of 5 spaced-apart SAW bursts acting in parallel.

### Leaky-mode SAW AOMs

As described elsewhere [8,9,18], a typical leaky-mode SAW AOM is an integrated optical device consisting at least of an in-coupling structure for light, an optical waveguide, and an interdigital transducer (IDT) [19] fabricated on a piezoelectric surface or substrate. Light enters the optical waveguide via the in-coupling structure, such as a prism pressed against the modulator surface in proximity to the waveguide, or an etched in-coupling grating [12]. The waveguide is defined by effective refractive indices ($n_{eff}$) for the guided modes, which are greater than the surrounding material refractive index ($n_{substrate}$) at the input polarization. The IDT induces a Rayleigh wave piezoelectric response at the modulator surface whose propagation speed is ≅3,600–4,000 m/s in our devices, depending on the SAW frequency. In our device, the SAW is counter-propagating to the optical waveguided mode(s). Where the waveguided light and SAW overlap, the SAW acts as a grating. This interaction has two impacts on a resulting optical signal: a portion is rotated to the orthogonal polarization, and it is diffracted into the bulk of the substrate as leaky-mode light [9]. As in Fig. 1, the angular deflection $\theta_{DIP}$ of the polarization-rotated optical signal within the substrate is determined by: the free-space optical wavelength, $n_{eff}$ of the guided mode in the waveguide, the SAW's temporal frequency component of interest ($f$), SAW velocity on the optical waveguide ($v_{SAW}$), and $n_{substrate}$ experienced by the polarization-rotated signal. $\theta_{DIP}$ is given by the following expression and depicted in Fig 1a: $\theta_{DIP} = \cos^{-1}[(k_{guided} + mk_{SAW}) / k_{signal}]$. The $k$-vectors in this expression are given by $k_{guided} = n_{eff} / \lambda_0$, $k_{SAW} = f / v_{SAW}$, and $k_{signal} = n_{substrate} / \lambda_0$, where $\lambda_0$ is the free space optical wavelength, $n_{substrate}$ is the refractive index at the signal polarization, and $m$ is the diffraction order, where $m$ = +1.

To illustrate the properties of a typical modulator, key quantities measured at 632.8 nm are provided. In *x*-cut, *y*-propagating LiNbO$_3$, TE-polarized input light (E-field along the *z*-axis) interacts with the extraordinary substrate index $n_e$ = 2.2022 and is guided into a desired optical waveguide mode such as $n_{eff, TE1}$, measured by a Metricon prism coupler to be 2.215 in our device. Driving the IDT from *f* = 250 – 400 MHz in the 5 mW regime induces SAWs having $v \cong$ 3,600–4,000 m/s along the waveguide, depending on *f*, optical waveguide type, and crystal cut. The TM-polarized (E-field in the *xy* plane) leaky-mode signal interacts with the ordinary substrate index $n_o$ = 2.2865. Referring to Fig. 1a, our surface-emitting devices operate at 640 nm and produce signals that travel at 5° < $\theta_{DIP}$ < 8° depending on waveguide mode and *f*.

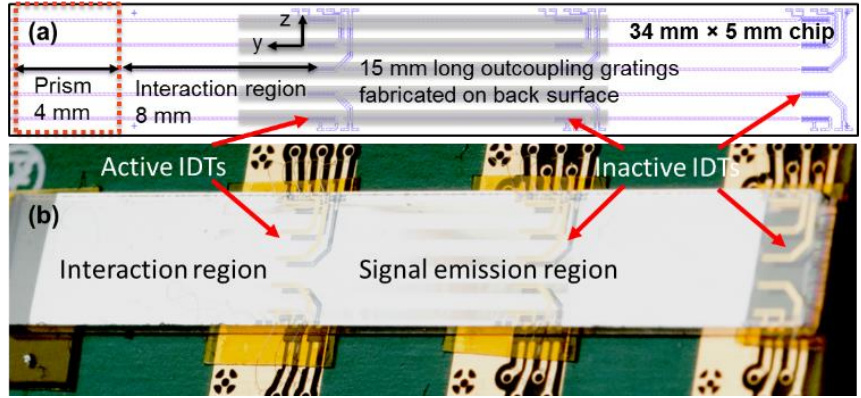

Fig. 2. (a) Device layout with aspects of the waveguides, IDTs, and backside out-coupling gratings shown. This experimental device has three columns of five IDTs. Only the first column was used in the scope of this report. (b) Device photograph.

## Modulator Design and Methods

The piezoelectric material chosen for the modulator of Fig. 2 is *x*-cut lithium niobate (LiNbO$_3$). Indiffused optical waveguides are created along the crystal *y*-axis, 100 μm wide along *z*, via annealed proton exchange (APE) followed by reverse proton exchange (RPE) to increase the index of refraction along the crystal's *z*-axis, enabling guiding of TE-polarized light propagating along the *y*-axis. Each waveguide has a corresponding IDT, typically measuring 620 μm along *y* in a chirped configuration with individual finger widths of 1.65–2.23 μm for broad RF response. The IDTs are patterned with a maskless aligner (Heidelberg Instruments MLA15) and deposited in Cr:Au. Background on IDTs is available in [19].

Light diffracted within a leaky-mode SAW modulator typically travels at a near-glancing angle to the waveguide, requiring an out-coupling grating or other angle-changing feature to overcome TIR if emission from a broad surface, rather than an edge, is desired. While volumetric gratings as in [16] are advantageous for near-eye augmented reality displays due to their partial transparency, they are unnecessary for desktop and mobile systems. Referring to Fig. 3, we used a backside surface grating for better compatibility with high-volume wafer processing. The grating was designed by a hill-climbing algorithm wrapping the S$^4$ Rigorous Coupled Wave Analysis (RCWA) package [20] with the metric of maximizing the diffracted power, with layout as shown in Fig. 3. The outgoing angle from the grating was set 12° off-normal to reduce reflection at the exit surface by the Brewster effect. The simulated efficiency of the as-fabricated geometry was 50%. The gratings were fabricated via e-beam lithography using spin-on glass resist (hydrogen silsesquioxane (HSQ)) paired with a charge dissipating agent.

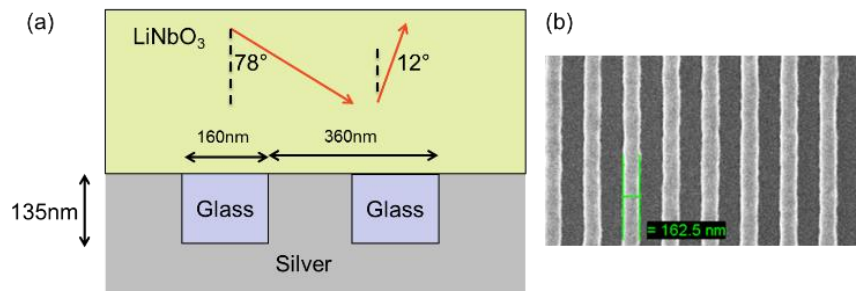

Fig. 3. (a) The out-coupling surface grating fabrication goal is 135 nm-thick spin-on HSQ with a 135–165 nm line width and 360 nm period, backed with silver. (b) SEM image of a test grating on LiNbO$_3$.

Fig. 4 shows the device test geometry. To predict the anticipated device behavior, we calculated $\theta_{DIP}$ for sequential drive signals from 290 to 320 MHz, the diffraction of the out-coupling grating, and exit refraction into air. An example "single-tone burst" waveform, e.g. Fig. 4b with tone duration 200 ns, induces a < 1 mm SAW at a primary frequency, e.g. *f*=290 MHz, should act as a grating that outputs diffracted light along a first trajectory. A series of similar sequential waveforms with tone bursts spanning the frequency range would cause output rays to incrementally scan at 0.01°/MHz in air, as in the orange (upper) line of Fig. 1c. Due to the reversed sign of our detector apparatus, defined in

Fig. 4a, the detector angle $\theta_{LAB}$ of the peak output optical power direction for each applied drive frequency $f$ will appear as a curve of negative rather than positive slope, with a constant angular offset, as will be discussed regarding Figs. 5-6.

Given the ~8 mm extent of the interaction region, between the prism and the IDT, the drive signal may be partitioned. In the example of signal 4c or 4d, four single-frequency bursts are delivered to the modulator, behaving as four spaced apart diffracting regions, which when illuminated produce four parallel, spaced-apart beams, with a trajectory controlled by the frequency.

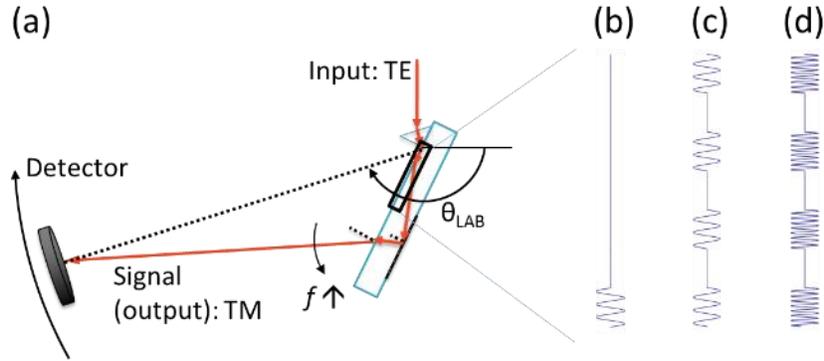

Fig. 4. (a) Top view of modulator characterization apparatus. The optical power detector is at the end of one of two rotating arms; observations are plotted with respect to arbitrary laboratory frame angle $\theta_{LAB}$. For increasing applied single frequency $f$, an output ray turns in the direction shown. (b, c, d) Example SAW waveforms.

## Results and Discussion

Our multi-channel AOM is mounted on a PCB and wire bonded. To induce SAWs, a 50Ω SMA jack on the PCB receives sinusoidal IDT drive waveforms of $f$ = 280–320 MHz from an HP 4648D signal generator, under computer control, via 28 dB and 12 dB gain stages. TE-polarized light from a 40 mW 640 nm laser diode is passed through a polarizer (not shown) and in-coupled using a rutile prism pressed to the front face of the modulator with coupling spot ~8 mm from the IDT. The prism-AOM-PCB assembly is placed on a manual rotational stage to excite the desired waveguide mode, which, here, is the $TE_1$-like mode. An HP 8130A dual pulse generator gates the computer-controlled IDT drive signal to the modulator channel in synchrony with a laser strobe signal.

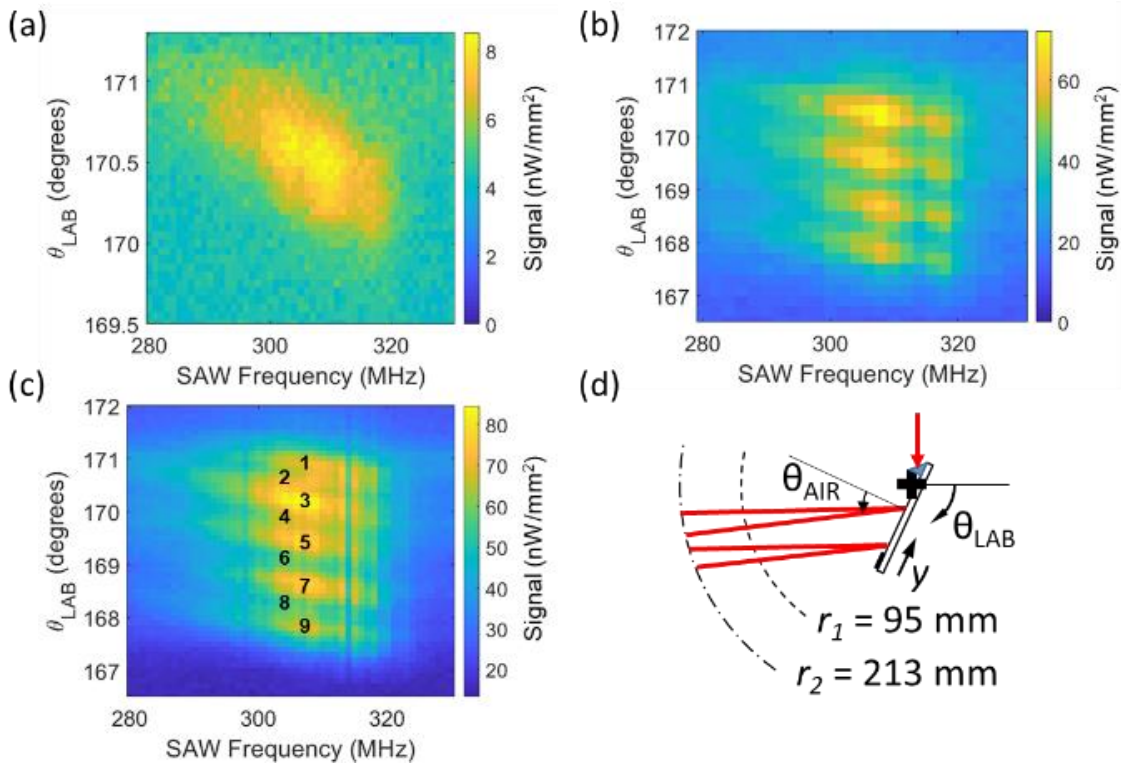

Fig. 5. (Color online.) Detected optical output power density, in units of nW/mm$^2$, as a function of applied waveform frequency and detector angle $\theta_{LAB}$. (a) A series of single-tone SAWs. (b) 4 spaced-apart SAW bursts act as 4 spaced-apart gratings. (c) Observation of beam steering from a 9-hogel electrohologram having 5 SAW fringes and 4 zero-amplitude spaces. (d) Emitted light exits the AOM at the frequency-dependent $\theta_{AIR}$, and is measured at two detector distances, $r_1$ = 95 mm and $r_2$ = 213 mm. Inverse ray tracing is used to determine the origin of the light along $y$ and its trajectory $\theta_{AIR}$ from $\theta_{LAB}$ and $r$.

In Fig. 4a, a TM-polarized output signal traverses the 1 mm-thick device, is redirected by the backside grating, exits the modulator surface, passes through a polarizer (not shown) to filter out unmodulated light, and is detected by a Thorlabs S130C slim photodiode detector behind an adjustable slit. To measure optical power output as functions of $f$ and detector angle $\theta_{LAB}$, the detector arm is incrementally rotated about an axis parallel to $z$ by a computer-controlled rotary stage and $f$ is ramped at each step. Henrie et al. [21] describe a linear version of a similar apparatus.

Motion of the SAW is implicitly frozen using short-pulse strobed light, an AOM illumination technique described in [13,22]. The pulse generator allows exploration of device behavior of various fringe lengths, spacings, and delays relative to the strobed illumination. The SAW traverses the 8-mm waveguide region between the prism and the IDT in 2.0-2.2 µs.

We first demonstrate the anticipated single beam emitter functionality of instances of SAW 1 of Fig. 1a spanning the circled operation points of the upper curve of Fig. 1c. RF and laser pulse widths are 300 ns, the SAW-positioning laser delay is 700 ns, and the repetition rate is 1 µs. For each waveform frequency, the resulting emitted signals are plotted in vertical cuts of Fig. 5a for a 40 mW peak / ~5 mW average laser power and 0.7 mm-width slit. The anticipated frequency dependence of each output signal's angle is observed in the negative slope of the plotted region, spanning ~0.8° in $\theta_{LAB}$ space, the angle recorded by the detector, from 290 to 320 MHz for detector distance $r_1$.

We next demonstrate that a SAW partitioned into hogels can occupy the device, enabling the emission of multiple discrete optical signals per Fig. 4(c, d). By setting the RF and laser pulse widths to 200 ns with a repetition period of 400 ns, i.e. less than the acoustic length of the modulator channel, four beams are emitted (vertical cut of Fig. 5b). Varying $f$ of the hogels is observed to change the detected angle of the beams' peak power, as expected, by regarding different vertical plot cuts of Fig 5c.

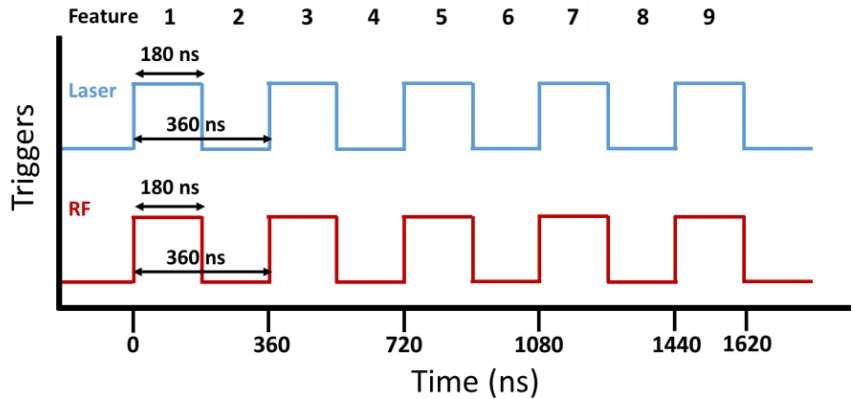

Fig. 6. Timing of the laser and RF waveform triggers corresponding to the 9-hogel datamap of Fig. 5c. In this example, there is no phase difference between the laser gating signal and the RF (IDT driver) gating signal.

To further explore the ability of arbitrary space-partitioning of the SAW interaction region, we decrease the RF and laser pulse widths to 180 ns and the period to 360 ns, resulting in the 5 beams corresponding to "time slots" for 9 single-frequency hogels in Fig. 5d. The pulse widths, period, offsets, and delay used to generate the 5-beam, 9-hogel case are shown in Fig. 6.

In a ray-optics approximation, our AOM output from a small surface patch has two degrees of freedom: origin along $y$ and trajectory in the $x$-$y$ plane. For increased analytical precision, these must be separated from optical peak power data in $\theta_{LAB}$ space by obtaining data at two sensor distances (95 mm and 213 mm) and performing inverse raytracing with knowledge of modulator orientation, as depicted in Fig. 5d. The experiment of Fig. 5b was run with the detector at two distances from the modulator. Through geometry and the values of $\theta_{LAB}$ of peak signal at $r_1$ and $r_2$, we find that for 294 < $f$ < 320 MHz, $\theta_{AIR}$ from hogel 3 spanned |29.2-29.5°| = 0.3°, corresponding to 0.3°/26 MHz = 0.01°/MHz, in agreement with prediction. The edge-to-edge extent of these 7 hogels spans approximately 8 mm along $y$, considered for a single $f$ to freeze the exit location. This result demonstrates that the entire length of the interaction region actively generates holographic fringes.

The duty cycle of the RF drive and laser strobe illumination is identical in this demonstration in order to improve the signal-to-noise ratio and determine the interaction length. In a display application, where different SAW frequencies are typically excited in each hogel, SAWs will be excited along the entire interaction length, followed a single pulse of strobe illumination. This is necessary to avoid mixing signal between different hogels, and is elaborated elsewhere [13, section 2.4]. Improved display brightness can be achieved in this case by utilizing a higher peak laser power to make up for the lower laser illumination duty cycle.

The IDTs utilized in these devices emit SAWs in both directions. No adverse effects due to SAW reflections are observed. The interaction between a SAW and the signal exiting the top surface of the modulator is expected to be negligible due to the short (µm-scale) interaction length. However, in a display, SAW absorbers and directional IDTs will be utilized to limit unwanted SAW propagation and reflections.

The modest angular output subtense of this first device can be expanded in several ways, such as configuration for a 100 MHz or larger operational bandwidth. The surface grating's broadband response is compatible with beam-steering via optical wavelength tuning, a future direction that would allow dramatically wider output angles alongside electronic drive [23].

In display applications, the SAWs will be composed of multiple frequency components and will be induced in modulator channels arrayed in two dimensions in each modulator device.

In this Letter, we described the application of a surface grating on the backside of a SAW modulator to provide a surface-emitting AOM, holding linear holograms of at least 8 mm that can be synchronously illuminated. Output light scanned at 0.01°/MHz in air for single- and multi-hogel waveforms.

## Acknowledgments

The authors acknowledge: A. Kopa, V. J. Bloomfield, D. A. Torres, M. Abban, M. M. Gleason, Y. W. Ho, and N. Orfanos for IDT design, modulator circuit board design, and assembly; L. Benney, S. Griffin, and A. Hare for the modulator characterization apparatus; M. G. Bancu for early microfabrication process development; and W. J. Shain for manuscript improvements. Some microfabrication was performed at the MIT MTL. Various technologies described here are patent pending.

# References Cited – Long Form